\documentclass[twocolumn,pre,superscriptaddress]{revtex4}

\usepackage[english]{babel}
\usepackage{amsmath}
\usepackage{amsthm}
\usepackage{enumitem}
\usepackage[usenames,dvipsnames]{xcolor}

\newlength{\extralength}
\usepackage{amssymb}
\usepackage[usenames,dvipsnames]{xcolor}
\usepackage{bbm}
\usepackage{cases}

\usepackage{float}
\usepackage{graphicx}
\usepackage{color}

\newcommand{\invisible}[1]{}
\newcommand{\bx}{\boldsymbol{x}}

\usepackage{amssymb}
\usepackage{times}
\usepackage{graphicx}
\usepackage[colorlinks=true, allcolors=MidnightBlue]{hyperref}

\usepackage{listings}

\begin{document}

\title{Dynamics of stochastic chains with harmonic and FPUT potentials}

\author{Emilio N. M. Cirillo}
\email{emilio.cirillo@uniroma1.it}
\affiliation{Dipartimento di Scienze di Base e Applicate per l'Ingegneria, 
             Sapienza Universit\`a di Roma, 
             via A.\ Scarpa 16, 00161, Roma, Italy.}

\author{Matteo Colangeli}
\email{matteo.colangeli1@univaq.it}
\affiliation{Dipartimento di Ingegneria e Scienze dell'Informazione e Matematica, Universit\`a degli Studi dell'Aquila, Via Vetoio, 67100 L'Aquila, Italy.}

\author{Claudio Giberti}
\email{claudio.giberti@univmore.it}
\affiliation{Department of Sciences and Methods for Engineering, University of Modena and
Reggio Emilia, via G. Amendola 2, 42122 Reggio Emilia, Italy.}
\affiliation{Interdepartmental Centers En\&Tech and InterMech, University of Modena and Reggio Emilia, Italy.}

\author{Lamberto Rondoni}
\email{lamberto.rondoni@polito.it}
\affiliation{Dipartimento di Scienze Matematiche, Politecnico di Torino, Corso Duca degli Abruzzi 24, 10129, Turin, Italy.}
\affiliation{INFN, Sezione di Torino,Via Pietro Giuria 1, 10125, Turin, Italy.}

\begin{abstract}

Inspired by recent studies on deterministic oscillator models, we introduce a stochastic one-dimensional model for a chain of interacting particles.
The model consists of $N$ particles performing continuous-time random walks on the integer lattice $\mathbb{Z}$ with exponentially distributed waiting times. 
The \textcolor{black}{particles} are bound by confining forces to two particles that do not move, placed at positions $x_0$ and $x_{N+1}$, respectively, and they feel the presence of baths with given inverse temperatures: $\beta_L$ to the left, $\beta_B$ in the middle, and $\beta_R$ to the right.
Each particle has an index and interacts with its nearest neighbors in index space through either a quadratic potential or a Fermi-Pasta-Ulam-Tsingou type coupling. This local interaction in index space can give rise to effective long-range interactions on the spatial lattice, depending on the instantaneous configuration. Particle hopping rates are governed either by the Metropolis rule or by a modified version that breaks detailed balance at the interfaces between regions with different baths.
In both cases, the dynamics drive the system toward the minimization of an appropriate energy functional, even under non-uniform temperature profiles.
\end{abstract}

\maketitle

%\renewcommand{\contentsname}{\textbf{Contents}\par}
%%\renewcommand{\contentsname}{\vskip -1.2 cm $\phantom.$\par}
%\addtocontents{toc}{\protect\setcounter{tocdepth}{-1}}
%\tableofcontents
%\addtocontents{toc}{\protect\setcounter{tocdepth}{3}}

%\newpage

\section{Introduction}
\label{sec:intro}

Chains of coupled oscillators are 
widely investigated models, 
as paradigmatic systems for studying the microscopic mechanisms underlying heat conduction \cite{rieder1967properties,Lepri2003,zhong2012normal,chen2016key,Dhar2008}. In their deterministic formulation, such chains consist of particles arranged on a one-dimensional lattice and interacting through prescribed forces. These models offer a minimal yet versatile framework for addressing fundamental questions, including the derivation of Fourier's law, 
\textcolor{black}{or of anomalous}
heat conduction from microscopic principles, 
the 
\textcolor{black}{properties}
of nonequilibrium steady states, and the rigorous formulation of hydrodynamic limits \cite{Dhar2008,Bernardin2005,SpohnLebowitz1977,CDP16,CDP17b,CGV23}. Depending on the 
interaction potential, deterministic oscillator chains 
display a rich spectrum of transport regimes, ranging from normal diffusion \cite{GLPV00} to anomalous heat conduction \cite{LW03,BLL20}.

An illustrative example is provided by the celebrated Fermi-Pasta-Ulam-Tsingou
(FPUT) model \cite{Fermi1955}, originally introduced to explore thermalization 
in nonlinear lattices. The FPUT chain revealed unexpected long-time correlations 
and slow relaxation, opening a research direction that connects ergodic theory, 
nonlinear dynamics, and statistical physics \cite{Ford92,Casetti1996,Gallavotti08,Benettin2018}.
Another cornerstone is the Toda model \cite{Toda1967}, which describes a 
chain of particles with exponential nearest-neighbor interactions. Unlike the 
FPUT chain, the Toda chain is integrable and supports soliton solutions. 
Together, the FPUT and Toda models exemplify the subtle interplay between 
nonlinearity, integrability, and transport properties in deterministic 
oscillator chains \cite{Ford92,benettin2013fermi,benettin2021fpu}.

Stochastic variants of oscillator chains have been introduced to simplify 
mathematical analysis and to model the effects of thermal fluctuations. 
Random noise can be incorporated either through stochastic perturbations 
of the equations of motion or by designing interacting particle systems whose 
invariant measures mimic those of oscillator chains \cite{KipnisLandim,Basile2010}. 
Stochastic models often preserve key conservation laws, such as energy or 
momentum, while breaking others.
However, their mathematical simplicity, compared to the complexity of 
deterministic dynamics, enables rigorous results on 
transport and scaling limits, thus shedding light also on properties of deterministic systems, that can only be studied numerically.
They hence provide testbeds for  studying the interplay between microscopic reversibility, detailed balance, and macroscopic irreversibility.
In particular, they may help understand properties of one-dimensional deterministic systems, \textcolor{black}{such as the breaking of local equilibrium and the onset of long-range correlations}, that appear to be peculiar because not found in usual thermodynamic systems, but still  
confirmed by accurate simulations, see {\em e.g.}\ Refs.\ \cite{hurtado2006breakdown,giberti2011anomalies,giberti2019temperature,giberti2019n,mejia2019heat,di2024microscopic}.

We investigate stochastic 
\invisible{oscillator}\textcolor{black}{particle} chains defined on a lattice 
(see, {\em e.g.}, \cite{CFPG2021} for an application to decohesion), 
whose dynamics follow either the Metropolis rule or a modified variant thereof, which breaks detailed balance 
\textcolor{black}{at the interface between
regions at different temperatures. In particular, there are three regions characterized by an
inverse temperature profile}
$\beta(x)$ on the lattice $\mathbb{Z}$,
\textcolor{black}{that is a step function
taking different values in the three regions, thus allowing the onset of nonequilibrium stationary states}.
The main objective is to examine the effect of 
\textcolor{black}{such}
spatially
varying temperature profiles, designed to produce stochastic
dynamics that are explicitly non-homogeneous in space. 
The resulting
\textcolor{black}{particles density}
profiles, obtained by calculating the average particle
positions and their mean jump intensities - which can be
interpreted as an analogue of kinetic energy in a continuous
model - are discussed in detail.

Remarkably, but analogously to deterministic $\beta$-FPUT chains under stretching (or negative pressure) conditions \cite{giberti2011anomalies,mejia2019heat,di2024microscopic},
the particles accumulate in the hottest
regions of the system, precisely at the sites where the
dynamics is fastest and particle mobility is greatest. The
analysis is performed using different interaction potentials,
and it is observed that the results depend strongly on this
choice, as expected. 
This shows 
how non-homogeneous stochastic dynamics 
can quite simply induce pronounced
spatial organization, and thermodynamically peculiar behaviors,
revealing numerous similarities with deterministic one-dimensional systems. The
accumulation of particles 
in regions of high temperature, already found in deterministic particle models, thus appears
to be of quite general validity. 

In Sec.~\ref{sec:sec1} we define the 
\invisible{oscillator}\textcolor{black}{particle} chain and introduce the mathematical framework. In 
Sec.~\ref{sec:sec2} we present numerical simulations for inhomogenous models, while conclusions are drawn in Sec.~\ref{sec:sec3}.

\section{The model}
\label{sec:sec1}
%We 
Consider a 
%stochastic 
model consisting of $N$ particles interacting on the integer lattice $\mathbb{Z}$, where $N$ is assumed to be odd. These particles, labeled $1, \ldots, N$, evolve in continuous time. Each particle performs random jumps on $\mathbb{Z}$ according to given transition rates. The interaction is local in index space, meaning that particle $k$ interacts only with particles $k-1$ and $k+1$. Hence, the model can be viewed as an 
interacting particle system in which individually labeled particles move on 
$\mathbb{Z}$ subject to nearest neighbor interactions in index space.

%By construction, the stochastic 
%dynamics satisfy detailed balance.

\invisible{The parameter $\beta$ of the Metropolis algorithm can be: 
   \begin{enumerate}
       \item {\bf \tt [model \tH]} independent of the labels of the particles;
       \item  {\bf \tt [model \text nH]}  dependent on the label of the particles: $\beta_k$ (with laws to be defined). 
   \end{enumerate}
   }
More precisely, let $\bx \in \mathbb{Z}^{N}$ denote the vector of particle 
positions $\{x_k\}_{k=1}^{N}$. In addition, we introduce two more particles with 
indices $k=0$ and $k=N+1$, fixed at the positions $x_0 = -a(N+1)/2$ and 
$x_{N+1} = a(N+1)/2$, where
$a$ is a positive integer.
These particles are 
pinned at prescribed sites of the lattice and remain fixed in time, thereby 
providing boundary conditions for the \invisible{oscillator}\textcolor{black}{particle} chain. 
The lattice $\mathbb{Z}$ is partitioned into three regions,
$x\le x_0$, $x_0<x<x_{N+1}$, and $x\ge x_{N+1}$, which are
respectively referred to as the \textit{left boundary}, the
\textit{bulk}, and the \textit{right boundary}.

The Hamiltonian of the system is given by:
\begin{equation}
    H(\bx) = \sum_{k=0}^{N} V(x_{k+1} - x_k),
    \label{eq:H}
\end{equation}
%where $V(r)$ denotes the interaction potential. 
for an interaction potential $V(r)$, which is
%In this work, the potential is taken to be 
either {\em harmonic}:
\begin{equation}
    V(r) = \frac{g_2}{2} (r - a)^2, 
    \quad g_2 > 0,
    \label{harm}
\end{equation}
or {\em Fermi--Pasta--Ulam--Tsingou} ($\beta$-FPUT) \cite{Fermi1955},
\begin{equation}
    V(r) = \frac{g_2}{2} (r - a)^2 
           + \frac{g_4}{4} (r - a)^4, 
    \quad g_2 > 0, \, g_4 > 0.
    \label{fpu}
\end{equation}
In both cases, the vector 
$\bx^e = (x_0 + a k)_{k=0,\ldots,N+1}$ 
identifies the mechanical equilibrium configuration of the system.
%of the mechanical system described by the Hamiltonian~\eqref{eq:H}.

It is worth noting that 
while the parameter $a$ %slightly 
modifies the form of the harmonic Hamiltonian only up to an additive constant, in the $\beta$-FPUT case it gives rise to a cubic term.
%, as follows:
%\begin{align}
%\label{harmHam}
%H(\bx)
%=-\frac{1}{2}g_2a^2(N+1)
%+\frac{1}{2}g_2
%\sum_{k=0}^{N}(x_{k+1}-x_k)^2
%\end{align}
%and
%\begin{align}
%\label{fpuHam}
%H(\bx)
%&
%=
%-\Big(
%\frac{3}{4}g_4a^4
%+\frac{1}{2}g_2a^2
%\Big)
%(N+1)
%+\Big(\frac{1}{2}g_2+\frac{3}{2}g_4a^2\Big)
%\sum_{k=0}^{N}(x_{k+1}-x_k)^2
%\notag\\
%&
%\phantom{=.}
%-g_4a
%\sum_{k=0}^{N}(x_{k+1}-x_k)^3
%+\frac{1}{4}g_4
%\sum_{k=0}^{N}(x_{k+1}-x_k)^4
%\end{align}
%respectively, in the harmonic and in the $\beta$-FPUT 
%case.

%\cgr{Ho (Claudio) quanche perplessità sulla parte precedente in rosso. Non c'è nulla di sbagliato, ma non è chiaro qule sia il ruolo di questa osservazione perché poi alla presenza dei termini di grado 3 in $x_{k+1}-x_k$ non si fa più riferimento. Inoltre mi sembra che dire che $a$ aggiunge un termine cubico a FPUT sia non proprio preciso, perché nella definizione stadard (o accettata come tale dalla gran parte degli autori) in FPU $a$ c'è (e duqnue c'è il termine cubico nella distanza fra le particelle).}

%\enmc{[Ho modificato la frase incriminata (vedi pezzo rimasto rosso). Ho rimosso add, che effettivamente era ingannevole, o l'ho riformulata cercando di mettere in luce che anche se l'equazione (3) che difinisce il potenziale non ha il termine cubico, in realta` nell'Hamiltoniana e` presente. Se vi dicessi che ho capito veramente il problema, mentirei, ma mi pare che quello che abbiamo scritto al momento non sia sbagliato.]}

The Hamiltonian \eqref{eq:H} is space homogeneous,
namely its form does not depend on the
position of the particles. We are interested in
studying models in which this homogeneity is broken,
in the sense that the transition rates of the dynamics
depend on the 
actual position occupied by the particles, and 
not simply on the relative position of the interacting particles.
We 
%will 
pursue this through two distinct Markov chains, defined
%\cgr{Mi pare che i transition rate dipendano dalla posizione occupata dalle particelle anche nel caso omogeneo, visto che dipendono da $x_{k+1}-x_k$. Forse qui si dovrebbe dire che, pur matenendo la dipendenza da $x_{k+1}-x_k$, si vuole rompere  l'invarianza per traslazione del potenziale che ci sarebbe ignorando le condizioni al contorno (o qualcosa del genere....)}
%\enmc{[Modificata anche questa (vedi frase rossa). Prima era scritta male, ma il senso era quello messo in luce da Claudio. Adesso mi pare risponda al problema sollevato da Claudio.]}
in terms of a
\textcolor{black}{three steps}
function
$\beta(x)$, with $x \in \mathbb{Z}$, which characterizes 
\textcolor{black}{three regions of the}
lattice. Fixing $\beta_0 \equiv \beta(0) > 0$,
%as a reference inverse temperature, evaluated for example at $x = 0$.
%Moreover, 
%for any $k=0,\dots,N$ 
we then define the \textit{effective potential} as
%\begin{equation}
%   \mathcal{E}(x_{k},x_{k+1})=\beta_0^{-1}\beta(x_k) ~ V(x_{k+1}-x_k),  \label{effpot}
%\end{equation}
\begin{equation}
\mathcal{V}(x,y)=\frac{1}{\beta_0}\tilde{\beta}(x,y) ~ V(y-x), \label{effpot} 
\end{equation}
where $\tilde{\beta}(x,y)=\left[\beta(x)+\beta(y)\right]/2$, for $x,y\in\mathbb{R}$.
%\begin{equation}
%\mathcal{V}(x_{k},x_{k+1})=\frac{1}{\beta_0}\tilde{\beta}(x_k,x_{k+1}) ~ V(x_{k+1}-x_k), \label{effpot} 
%\end{equation}
%where $\tilde{\beta}(x_k,x_{k+1})=\left[\beta(x_k)+\beta(x_{k+1})\right]/2$.
While $V$ depends only on the distance %position 
of two particles, 
%with consecutive indices, 
the effective potential $\mathcal{V}$ depends
on the positions of both, hence it is not
%particles.
%, also, 
%while $\tilde\beta$ is 
symmetric in the
exchange of its arguments, 
%$\cal V$
%the effective potential 
%is not 
\textcolor{black}{if the two particles lie in regions with different temperatures}. 
The %corresponding 
\textit{effective energy}
%or \textit{Hamiltonian}
of the \invisible{oscillator}\textcolor{black}{particle chain} is then defined by
\begin{equation}
\mathcal{H}(\bx) = \sum_{k = 0}^{N} \mathcal{V}(x_k, x_{k+1}),
\label{effen}
\end{equation}
which, as $H(\bx)$ in Eq.~\eqref{eq:H}, also attains its global minimum at $\bx^e$. 

\textcolor{black}{Using these notions, in the next sections we introduce two different stochastic dynamics that, for brevity, 
are respectively called {\em reversible} or
{\em irreversible}, depending on whether the condition of detailed balance is fulfilled or not.}\footnote{This terminology is customary in the stochastic processes approach to problems in statistical mechanics, because systems satisfying detailed balance allow events as well as their time reverses, and also because they are commonly assumed to arise from underlying time-reversal-invariant microscopic dynamics, see e.g. \cite{kreuzer,wehrl,Presutti08,Spohn1991}. Naturally, the stochastic notion of reversibility is quite different from that of microscopic dynamics, whether classical or quantum \cite{sachs,robnik,lebowitzRMP,carbone,Chamber24}. Indeed, some authors take stochastic processes as fundamental, with no need to refer to any lower-level description \cite{streater,kardar}. Even when the stochastic description is regarded as a coarse graining of lower-level dynamics, microscopic reversibility is neither necessary nor sufficient for detailed balance to hold \cite{jepps2016,dal,klages,colan12}.}

\subsection{Reversible dynamics}
\label{sec:sec1-rev}
The 
%stochastic 
continuous time dynamics of the \textcolor{black}{particle system} is defined as follows. Let $\bx$ be the configuration of the system at time $t$. 
At a later time $t + \tau$, where $\tau$ denotes 
a random time increment, the configuration $\bx$ changes to a new one, $\bx^{(j,\pm)} = (x_1, \dots, x_{j-1}, x_j \pm 1, x_{j+1}, \dots, x_N)$, with $j \in \{1, \ldots, N\}$. In this new configuration, a randomly selected particle, say the $j$th, has moved one step to the right ($+$) or to the left ($-$), 
%on the lattice, 
reaching position $ x_j \pm 1$.
%Correspondingly, we denote $\mathcal{E}(\bx'^{(j,\pm)})=\sum_{k=0}^N\mathcal{E}(x'_{k},x'_{k+1})$
%$\mathcal{E}'_k\equiv \mathcal{E}(x'_{k+1}-x'_{k})$ and 
%$\mathcal{E}'^{(j,\pm)}\equiv\mathcal{E}(\bx'^{(j,\pm)})$. 
We consider the Markov jump process with the Metropolis rates \cite{Madras2002}
\begin{equation}
c(\bx,\bx^{(j,\pm)})=
           \min \left\{1, e^{-\beta_0 \Delta^{(j,\pm)} \mathcal{H}(\bx)}\right\} ,
%           \min \left\{1, e^{-\beta_o\left[\mathcal{E}(x_{j-1},x_{j}\pm 1)+\mathcal{E}(x_{j}\pm 1,x_{j+1})-\mathcal{E}(x_{j-1},x_{j})-\mathcal{E}(x_{j},x_{j+1})\right]}\right\} .
      \label{eq:intens_1}
\end{equation}
associated with transitions from the state $\bx$ to the state $\bx^{(j,\pm)}$, 
where $\Delta^{(j,\pm)} \mathcal{H}(\bx)=\mathcal{H}(\bx^{(i,\pm)})-\mathcal{H}(\bx)$.
%\begin{align*}
%\Delta^{(j,\pm)} \mathcal{H}(\bx)
%&=
%\mathcal{H}(\bx^{(i,\pm)})
%-
%\mathcal{H}(\bx)
%\nonumber\\
%&=
%\mathcal{V}(x_{j-1},x_{j}\pm1)
%+\mathcal{V}(x_{j}\pm1,x_{j+1})
%-[\mathcal{V}(x_{j-1},x_{j})+\mathcal{V}(x_{j},x_{j+1})]
%\nonumber\\
%&=
%\tilde{\beta}(x_{j-1},x_{j}\pm1)V(x_{j}\pm1-x_{j-1})
%\notag\\
%&
%\phantom{=}
%+\tilde{\beta}(x_j\pm1,x_{j+1})V(x_{j+1}-(x_{j}\pm1))
%\nonumber\\
%&
%\phantom{=}
%-[\tilde{\beta}(x_{j-1},x_{j})V(x_{j}-x_{j-1})
%+\tilde{\beta}(x_j,x_{j+1})V(x_{j+1}-x_{j})].
%\end{align*} 
Therefore, the jump intensity of the $j$th particle reads 
\begin{equation}
      c_{j}(\bx)=c(\bx,\bx^{(j,+)})+c(\bx,\bx^{(j,-)}).
%      c_{j}(\bx)=c_{j}^{-}(\bx)+c_{j}^{+}(\bx).
\end{equation}

We consider the probability distribution 
$\pi(\bx)=e^{-\beta_0 \mathcal{H}(\bx)}/Z$, with $Z>0$ \textcolor{black}{the} normalization constant.
It is straightforward to verify that the Markov chain defined in \eqref{eq:intens_2} is reversible with respect to $\pi(\bx)$, 
\textcolor{black}{in the usual sense for stochastic processes, i.e. that }
the condition of detailed balance holds:
\begin{equation}
    \pi(\bx) c(\bx,\bx^{(j,\pm)}) 
    =\pi(\bx^{(j,\pm)}) c(\bx^{(j,\pm)},\bx),
    \label{detbal}
\end{equation}
since,
%\begin{align*}
%\pi(\bx) c(\bx,\bx^{(j,\pm)}) 
%&
%=
%\frac{1}{Z}e^{-\beta_0 \mathcal{H}(\bx)}
%\min 
%\Big(
%1, 
%e^{-\beta_0[\mathcal{H}(\bx^{(j,\pm)})- \mathcal{H}(\bx)]}
%\Big)
%\\
%&
%=
%\frac{1}{Z}
%\min 
%\Big(
%e^{-\beta_0 \mathcal{H}(\bx)}, 
%e^{-\beta_0\mathcal{H}(\bx^{(j,\pm)})}
%\Big),
%\end{align*}
%\begin{equation*}
%\pi(\bx) c(\bx,\bx^{(j,\pm)}) 
%=
%\pi(\bx)
%\Big(
%1 \wedge 
%e^{-\beta_0\Delta^{(j,\pm)} \mathcal{H}(\bx)}
%\Big)
%=
%\frac{1}{Z}
%\Big(
%e^{-\beta_0 \mathcal{H}(\bx)} \wedge 
%e^{-\beta_0\mathcal{H}(\bx^{(j,\pm)})}
%\Big),
%\end{equation*}
%which obviously coincides with 
%$\pi(\bx^{(j,\pm)}) %c(\bx^{(j,\pm)},\bx)$.
\begin{eqnarray}
\pi(\bx) c(\bx,\bx^{(j,\pm)}) 
&=&
\frac{1}{Z}
\min\Big(
e^{-\beta_0 \mathcal{H}(\bx)}, 
e^{-\beta_0\mathcal{H}(\bx^{(j,\pm)})}
\Big)\nonumber\\
&=&\pi(\bx^{(j,\pm)}) c(\bx^{(j,\pm)},\bx).
\end{eqnarray}
The reversibility property expressed by \eqref{detbal} also implies that $\pi(\bx)$ is the invariant distribution for the considered Metropolis dynamics \cite{Madras2002}.
Thus, at stationarity the dynamics 
samples states from $\pi(\bx)$. 
The Markov process is constructed by selecting 
the particle with index $j \in \{1,\ldots, N\}$ to be moved according to the distribution 
$c_{j}(\bx)/\sum_{k=1}^N c_{k}(\bx)$
and by allowing it to jump
backward with probability $c(\bx,\bx^{(j,-)})/c_j(\bx)$ and forward with probability $c(\bx,\bx^{(j,+)})/c_j(\bx)$.
Finally, the random 
\textcolor{black}{time}
increment $\tau$ is an exponentially distributed random variable with mean value $1/\sum_{k=1}^N c_{k}(\bx)$.

%As mentioned above, 
%In the inhomogeneous case,
%is
%introduced to mimic a situation in which 
%particles
%are coupled to different thermal baths along the lattice.
%In this interpretation, 
%$\tilde\beta/\beta_0$ represents
%the spatial modulation of the temperature.
%An
%alternative
%reading of this
%\cgr{[o preceeding?]} 
%\enmc{[messo un neutro e pilatesco this]}
%discussion is to
%assume 
%\textcolor{black}{view is} that the 
%\textcolor{black}{baths temperature is uniform} along the lattice
%\textcolor{black}{and takes the value}
%by the parameter 
%$\beta_0$, while $\tilde\beta$ tunes the interaction strength in the Hamiltonian, as if
%, depending on the spatial position, 
%the particles
%exerted forces on each other 
%\textcolor{black}{that depend on the region  they occupy, and not just on their distance or their relative positions.}
%with softened or
%hardened intensity.

\subsection{Irreversible dynamics}
\label{sec:sec1-nonrev}
Here, we compare the dynamics defined by Eq.~\eqref{eq:intens_1}, which is reversible with respect to $\pi$, with the alternative dynamics defined by the rates
%\begin{equation}
%      c(\bx,\bx'^{(j,\pm)})=\min \{1, e^{-\beta(x_j) \Delta_j^{(\pm)} H(\bx)}\},
%      \label{eq:intens}
%\end{equation}
\begin{equation}
c(\bx,\bx^{(j,\pm)})=
           \min \left\{1, e^{-\beta(x_j)\Delta^{(j,\pm)} H(\bx)}\right\} ,
      \label{eq:intens_2} 
\end{equation}
where $\Delta^{(j,\pm)} H(\bx)=
H(\bx^{(j,\pm)})-H(\bx)$.
%\begin{align*}
%\Delta^{(j,\pm)} H(\bx)
%&=
%H(\bx^{(j,\pm)})
%-H(\bx)
%\\
%&=
%V(x_{j}\pm1-x_{j-1})+V(x_{j+1}-(x_{j}\pm1))
%\\
%&
%\phantom{=}
%-[V(x_{j}-x_{j-1})+V(x_{j+1}-x_{j})]. 
%\end{align*}
This dynamics %does not satisfy Eq.~\eqref{detbal}, and 
is referred to %hereafter 
as {\it irreversible}, in the sense that it does not fulfill the detailed balance condition expressed by Eq.~\eqref{detbal}.
%\textcolor{black}{in the sense that it does not verify detailed balance} with respect to $\pi(\bx)$.
%\textcolor{black}{
%In our numerical experiments, beside evaluating %
%Eq.~\eqref{eq:intens_2}, we also replaced the term $\beta(x_j)\Delta^{(j,\pm)} H(\bx)$ appearing in the exponential of Eq.~\eqref{eq:intens_2} and expressed by the formula {\it IRR1} with the following expression:
%\begin{eqnarray*}
%\tilde{\beta}(x_{j-1},x_{j})V(x'_{j}-x_{j-%1})+\tilde{\beta}(x_j,x_{j+1})V(x_{j+1}-
%x'_{j})\nonumber\\
%-\tilde{\beta}(x_{j-1},x_{j})V(x_{j}-x_{j-1})-\tilde{\beta}(x_j,x_{j+1})V(x_{j+1}-x_{j}), \qquad (IRR2)
%\end{eqnarray*} 
%which lead to a distinct irreversible dynamics.}

The two dynamics defined by
Eqs.~\eqref{eq:intens_1} and \eqref{eq:intens_2} exhibit distinct features.
%are
%not equivalent. 
In particular, the Metropolis rule in
Eq.~\eqref{eq:intens_1} evaluates the parameter
$\beta(x)$ at both the updated position $x_j \pm 1$ and
the original position $x_j$ of the hopping particle,
assuming that a particle \textcolor{black}{immediately} thermalizes with its arrival position.
Differently,
the variant in Eq.~\eqref{eq:intens_2} depends
only on $\beta(x_j)$. 
\textcolor{black}{That means that the particle is considered at thermal equilibrium with the starting
position, but it does not immediately thermalize with the arrival one.}
With the latter dynamics, particles
evolve according to the space-homogeneous Hamiltonian
$H$, but with rates that are not spatially homogeneous
due to the temperature modulation encoded in
$\beta(x)$.
\textcolor{black}{On the other hand, 
the difference between the 
``reversible'' and the ``irreversible'' dynamics plays a role only at the interfaces
between two regions with different temperatures: namely at the two interfaces between the bulk region in which the system mostly lives, and the infinite left and right heat baths. As often the case, the irreversibility only concerns the coupling with the heat baths, and that produces transport phenomena.}

We 
%conclude this section with a few remarks, %providing a
%brief recap of the main properties of the two dynamics
\textcolor{black}{note} that 
%we have considered. First, 
while the particles with labels $0$ and $N+1$ are pinned on the lattice, the moving particles with indices $k \in \{1, \ldots, N\}$ are not confined to the \textcolor{black}{region of the lattice comprised between $x_0$ and 
$x_{N+1}$}, 
\textcolor{black}{and we have two scenarios:}
%We also underline that the system does not obey an exclusion principle, because multiple particles can occupy the same site of the lattice at a given time.
%\footnote{{We observe, however, that the exclusion principle could be introduced through a suitable potential with a repulsive component. For instance: $V_0(r)=\mbox{const.}$ for $r>0$ and $V(0)=\infty$ (in this case the model reproduces the classical SEP) or a soft-point potential $V(r)= (r^2+r^{-2})/2$ that introduces a smooth repulsive component alongside an attractive harmonic one.}}.
%
%Moreover, the definition of the function $\beta(x)$, with $x \in \mathbb{Z}$, can give rise to different physical scenarios:
\begin{enumerate}
\item \emph{Homogeneous models.} When $\beta(x) = \beta_0$ is constant across the lattice, the function $\mathcal{H}(\bx)$ reduces to $H(\bx)$, and %the invariant distribution 
$\pi(\bx)$ reproduces the standard Gibbs measure of equilibrium statistical mechanics \cite{Presutti08}. In this case, the two dynamics defined by Eqs.~\eqref{eq:intens_1} and \eqref{eq:intens_2} coincide.
\item \emph{Inhomogeneous models.} When $\beta(x)$ varies 
\textcolor{black}{in passing from one region to another of the three
constituting}
%along 
the lattice, the function $\mathcal{H}(\bx)$ no longer coincides with $H(\bx)$. In this setting, the two dynamics defined by Eqs.~\eqref{eq:intens_1} and \eqref{eq:intens_2} yield distinct stationary states, whose properties depend sensitively on the chosen interaction potential.
\end{enumerate}

The analysis of the inhomogeneous models %introduced above
is given in Sec.~\ref{sec:sec2}, with the
results of extensive numerical simulations of  harmonic and $\beta$-FPUT potentials,
under the dynamics defined in Eqs.~\eqref{eq:intens_1} and
\eqref{eq:intens_2}. The spatial
profiles of the mean particle positions and of the average jump intensities are examined.

%Observables:
%\begin{enumerate}
%    \item $\langle x_{k} \rangle, \quad k=1,\ldots, N$.
%    \item  $\langle x_{k+1} - x_{k} \rangle, \quad k=1,\ldots, N-1$.
%    \item $\langle \bar{x} \rangle$, where $\bar{x} = \frac 1N \sum_{k=1}^N x_k, k=1,\ldots, N$.
%    \item Let $\Delta_k^+ H$ the energy difference between two states $(x_j)_j$ and $(x_j+\delta_{j,k})_j$ and $\Delta_k^- H$ the energy difference between two states $(x_j)_j$ and $(x_j-\delta_{j,k})_j$. We consider the intensity, at each time step,
%    \begin{equation}
%        c(k)=\min \{ e^{-\beta_k \Delta_k^+ H},1\} + \min \{ e^{-\beta_k \Delta_k^- H},1\}, 
%    \end{equation}
%    and compute $\langle c(k)\rangle, k=1,\ldots, N$.
% \end{enumerate}

%\endnote{This is an endnote.} % To use endnotes, please un-comment \printendnotes below (before References). Only journal Laws uses \footnote.

% The order of the section titles is different for some journals. Please refer to the "Instructions for Authors? on the journal homepage.
%%%%%%%%%%%%%%%%%%%%%%%%%%%%%%%%%%%%%%%%%%

\begin{figure}[t]
\centering
{\includegraphics[width=0.6 \linewidth]{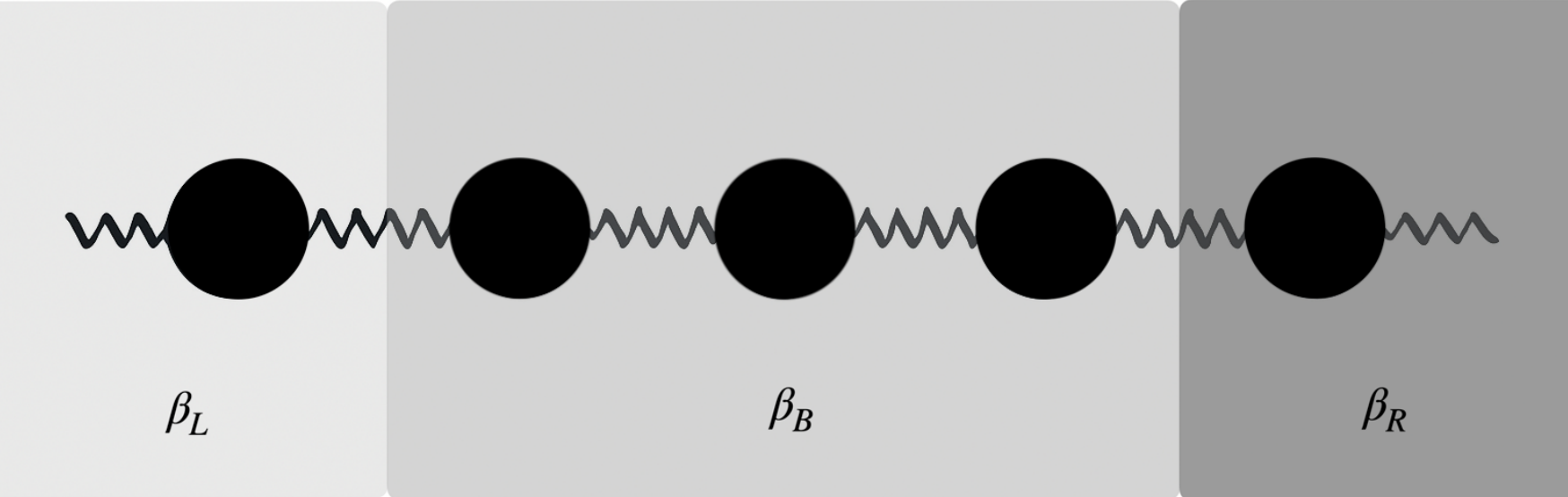}}
\caption{\textcolor{black}{The particle} chain considered in Sec.~\ref{sec:sec2} with $N=5$. Particles interact via either a harmonic or a $\beta$-FPUT potential and are in contact with three external thermal reservoirs, which fix the inverse temperature at $\beta_L$ on the left, $\beta_B$ in the center and $\beta_R$ on the right.}
  \label{fig:model}
\end{figure}

\section{Numerical simulations}
\label{sec:sec2}
\textcolor{black}{Consider the inhomogeneous model
with $\beta(x)$ a piecewise constant 
function, determined by the parameter 
$\delta \in [0, (N+1)a/2]$, so that:  $\beta(x) = \beta_L$, if $x \leq x_0 + \delta$; $\beta(x) = \beta_R$, if $x \geq x_{N+1} - \delta$; and $\beta(x) = \beta_B$,
if $x \in (x_0 + \delta , x_{N+1} - \delta)$,
with $\beta_L$, $\beta_R$ and $\beta_B$ are fixed positive parameters.}
This yields a \textcolor{black}{particle chain} coupled to three
thermal reservoirs,
\textcolor{black}{whose size depends on $N$ and $\delta$, that affect a particle's dynamics when it enters in the corresponding region of space,}
%
%. The first reservoir fixes the inverse
%temperature $\beta_L$ for the fraction of moving particles
%located in the left boundary and in a portion of the bulk on
%the left, whose size depends on the value of $\delta$. 
%Similarly, a second reservoir fixes the inverse temperature
%$\beta_R$ for the fraction of moving particles located in the
%right boundary and in a portion of the bulk on the right,
%whose size depends on the value of $\delta$. Finally, the
%third reservoir fixes the inverse temperature $\beta_B$ for
%particles located in the central part of the bulk between
%$x_0+\delta$ and $x_{N+1}-\delta$, 
see Fig.~\ref{fig:model}.

In this formulation, the parameter $\delta$ accounts for the presence of thermal layers near the endpoints of the chain, placed at $x_0$ and $x_{N+1}$. For a given interaction potential, smaller values of $\beta(x)$ correspond to higher hopping intensities for particles departing from site $x$. Conversely, for a fixed $\beta(x)$, the specific form of the interaction potential significantly influences the hopping rates, as shown in Eqs.~\eqref{eq:intens_1} and \eqref{eq:intens_2}. Therefore, the average intensity profile depends both on the form of the interaction potential and on the values of the parameters $\beta_L$, $\beta_B$, and $\beta_R$.

%\begin{figure}
%\centering
%{\includegraphics[width=0.49 \linewidth]%{Figs/fig1a-harm-eq.pdf}}
%\hfill
%{\includegraphics[width=0.49 \linewidth]{Figs/fig1b-harm-eq.pdf}}
%\caption{Normalized average position profile (left panel) and average intensity profile (right panel) in the homogeneous case with $\beta_L = \beta_B = \beta_R = 0.03$, $N = 21$ and $a = 10$.
%Solid disks and empty squares correspond, respectively, to the Metropolis dynamics defined in Eq.~\eqref{eq:intens_1} and to its variant given in Eq.~\eqref{eq:intens_2}.
%Simulations are performed over $5\times 10^7$ time steps using a harmonic potential.}
%  \label{fig:fig1}
%\end{figure}
 
\begin{figure}[t]
\centering
{\includegraphics[width=0.49 \linewidth]{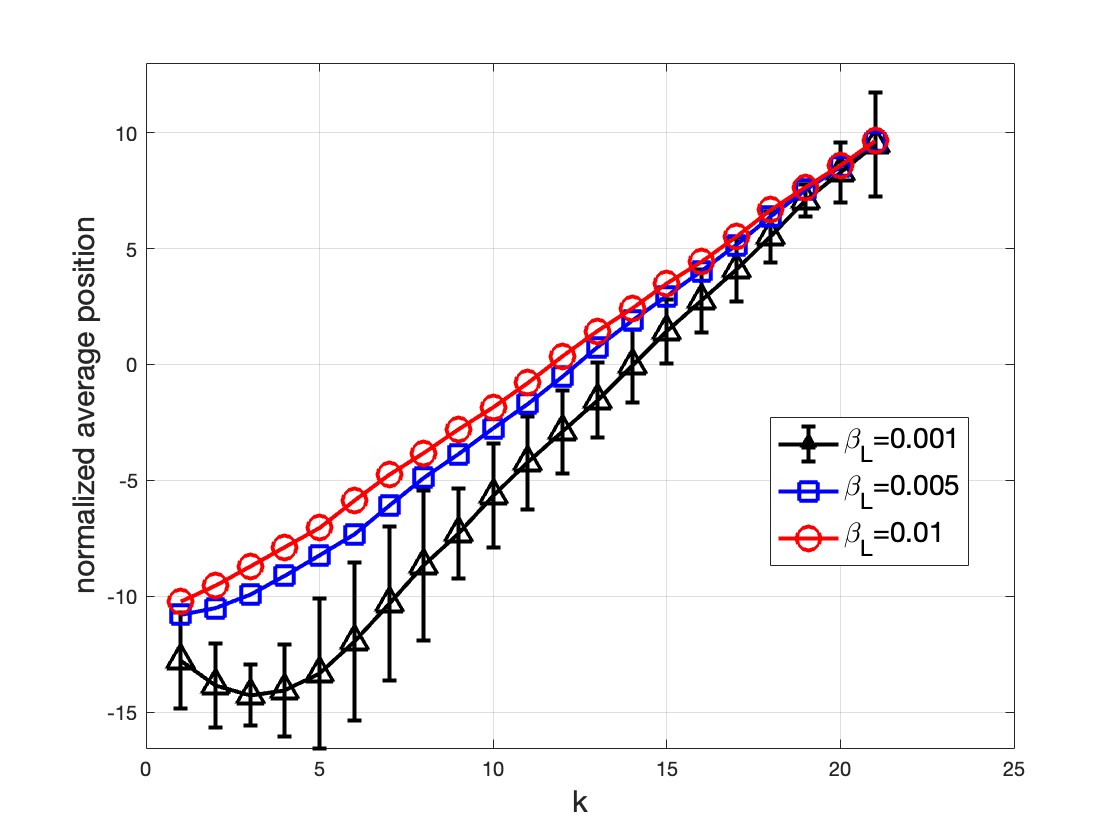}}
\hfill
{\includegraphics[width=0.49 \linewidth]{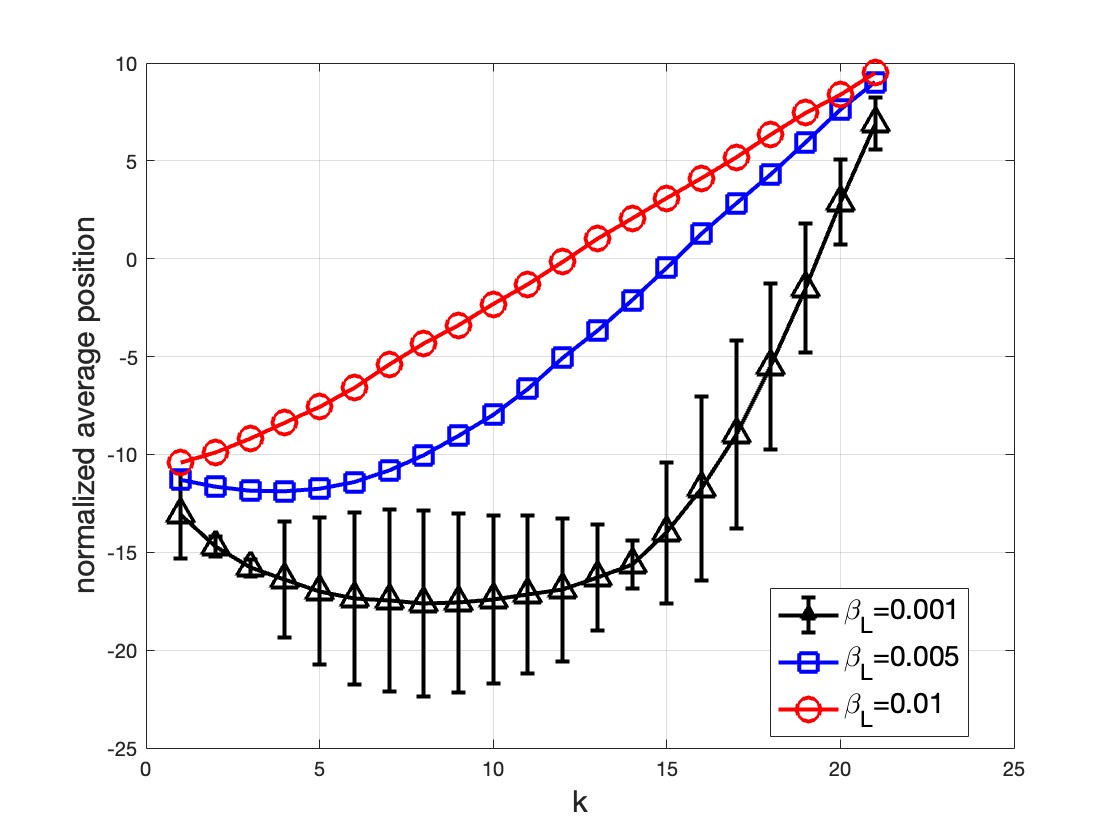}}
{\includegraphics[width=0.49 \linewidth]{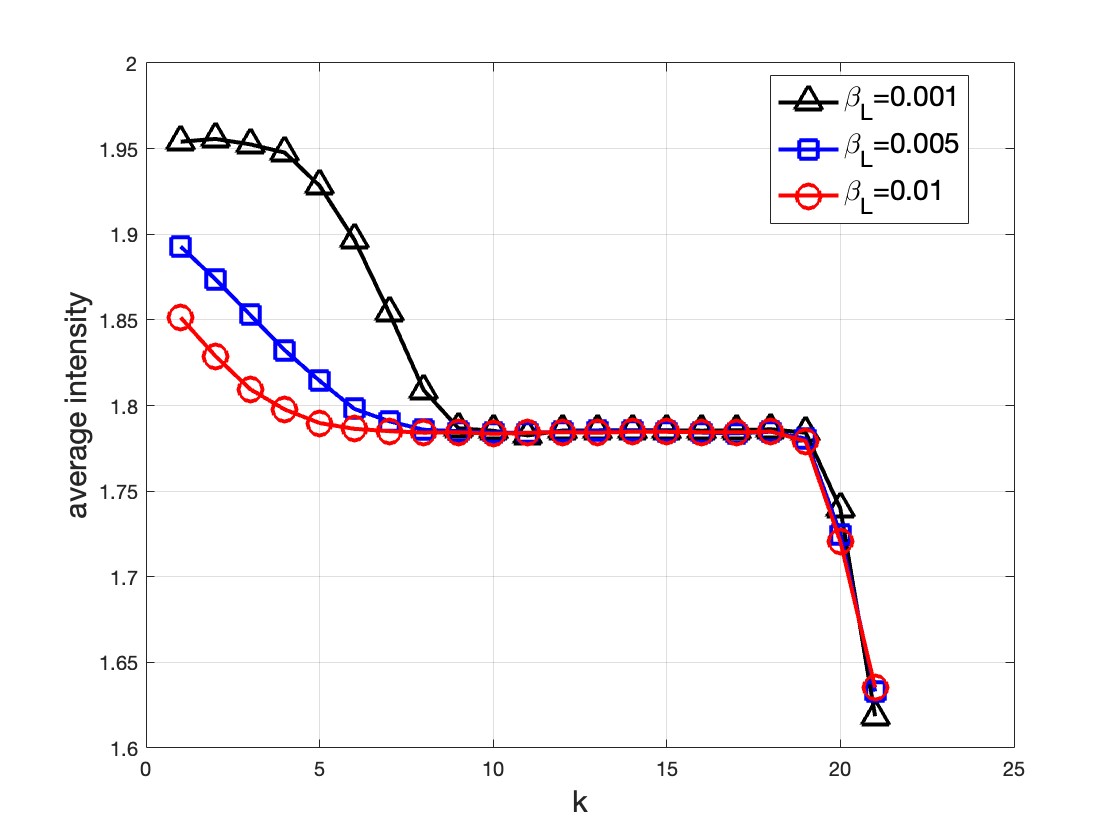}}
\hfill
{\includegraphics[width=0.49 \linewidth]{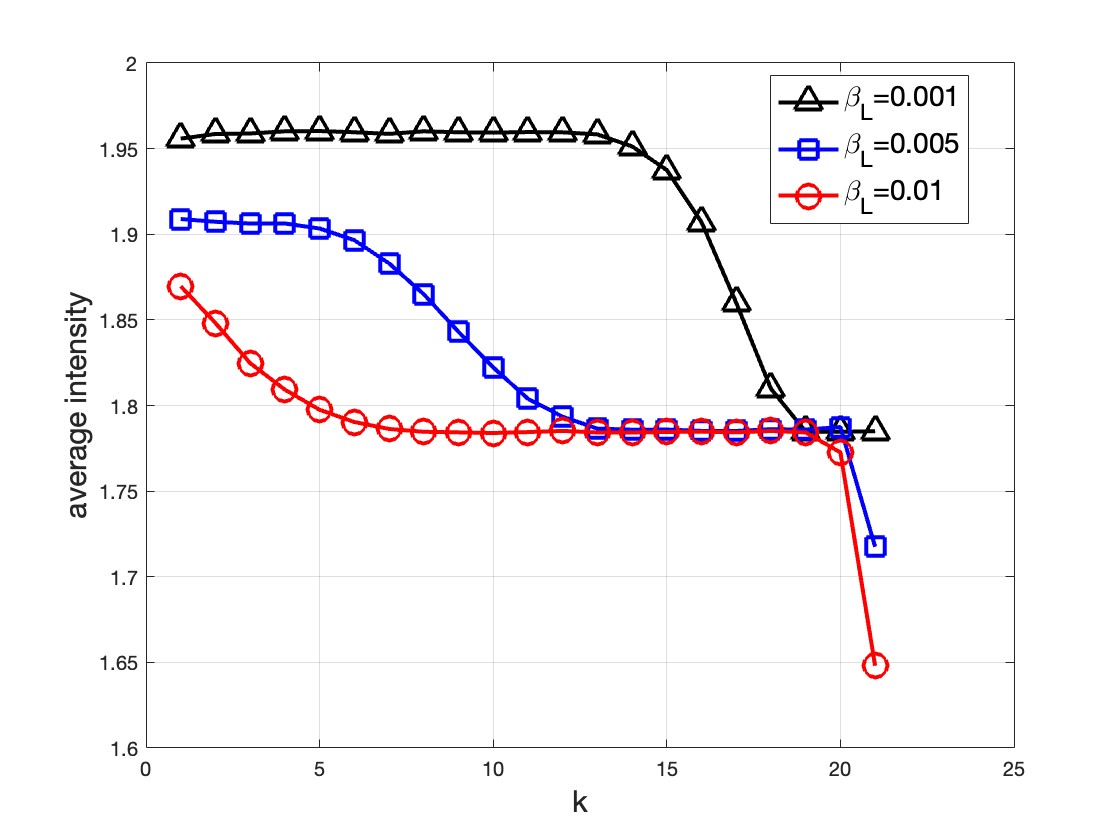}}
\vspace{-0.5cm}
\caption{\textit{Top row}: normalized average position profiles obtained with the harmonic potential in Eq.~\eqref{harm}, using the Metropolis dynamics defined in Eq.~\eqref{eq:intens_1} (left panel) and its variant given in Eq.~\eqref{eq:intens_2} (right panel), in the inhomogeneous case with $\beta_L\in\{0.001,0.005,0.01\}$, $\beta_B = 0.03$ and $\beta_R = 0.1$, with $N = 21$, $a = 10$, $\delta=1.5\ a$ and $g_2=1$. The errorbars represent one standard deviation. 
\textit{Bottom row}: average intensity profiles corresponding to the Metropolis dynamics (left panel) and to  Eq.~\eqref{eq:intens_2} (right panel).
Simulations are performed over $5\times10^7$ time steps.
\textcolor{black}{Various forms of contact resistance arise in all cases, analogously to
common deterministic models.}}
  \label{fig:fig2}
\end{figure}

%\begin{figure}
%\centering
%{\includegraphics[width=0.49 \linewidth]{Figs/fig2a-harm-d1.pdf}}
%\hfill
%{\includegraphics[width=0.49 \linewidth]{Figs/fig2b-harm-d1.pdf}}
%\caption{Average intensity profiles obtained with the harmonic potential using the Metropolis dynamics defined in Eq.~\eqref{eq:intens_1} (left panel) and its variant given in Eq.~\eqref{eq:intens_2} (right panel), in the non-homogeneous case with $\beta_L\in\{0.001,0.005,0.01\}$, $\beta_B = 0.03$ and $\beta_R = 0.1$, with $N = 21$, $a = 10$ and $\delta=1.5\ a$.
%Simulations are performed over $5\times10^7$ time steps.}
%  \label{fig:fig3}
%\end{figure}

\begin{figure}[t]
\centering
{\includegraphics[width=0.49 \linewidth]{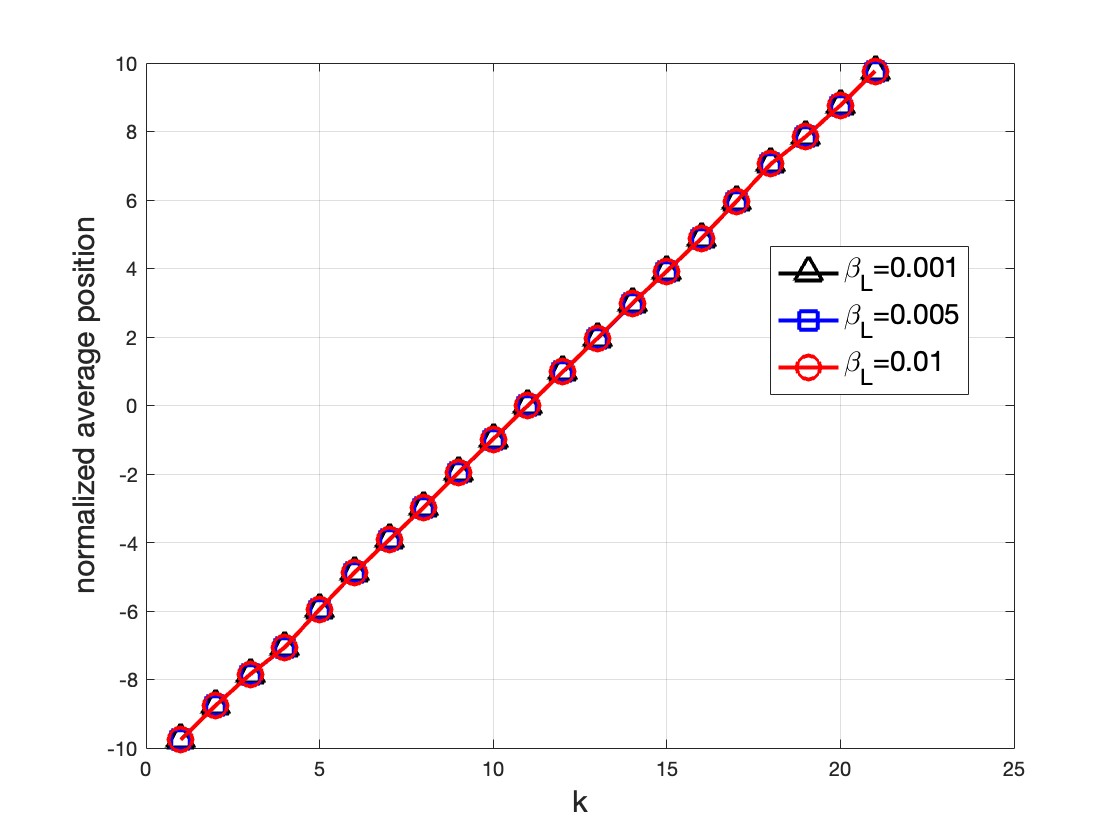}}
\hfill
{\includegraphics[width=0.49 \linewidth]{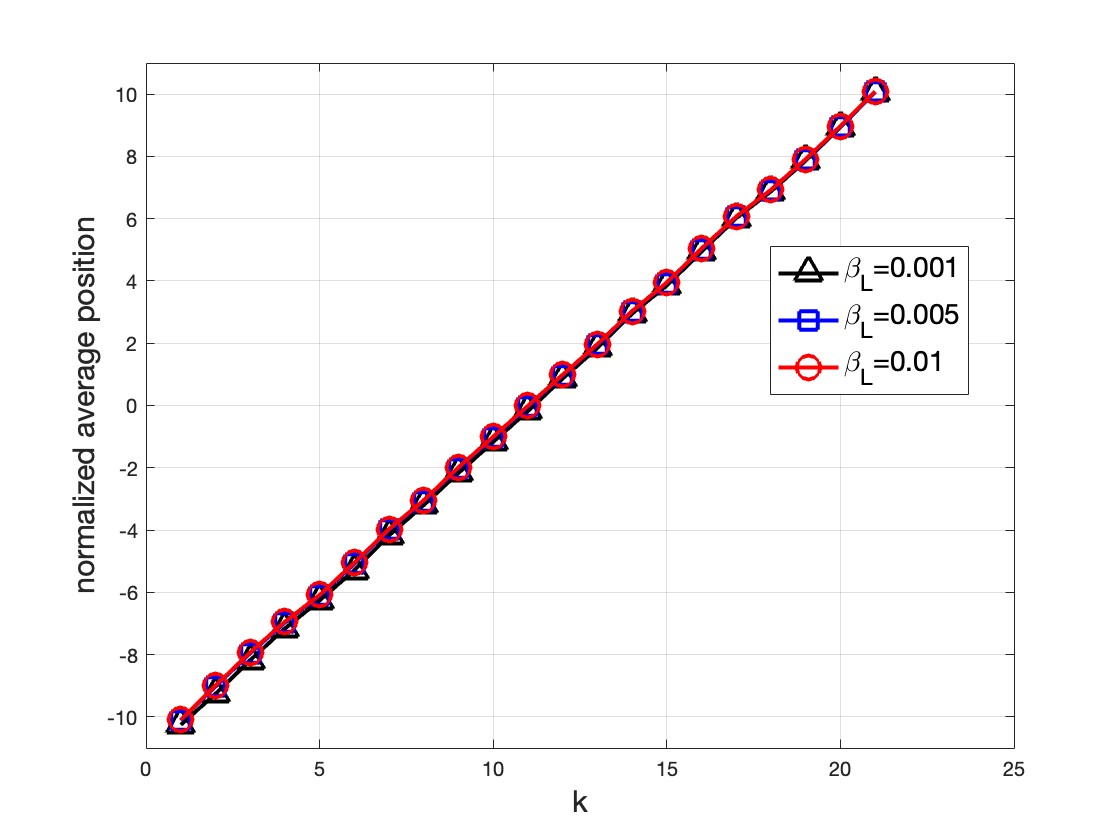}}
{\includegraphics[width=0.49 \linewidth]{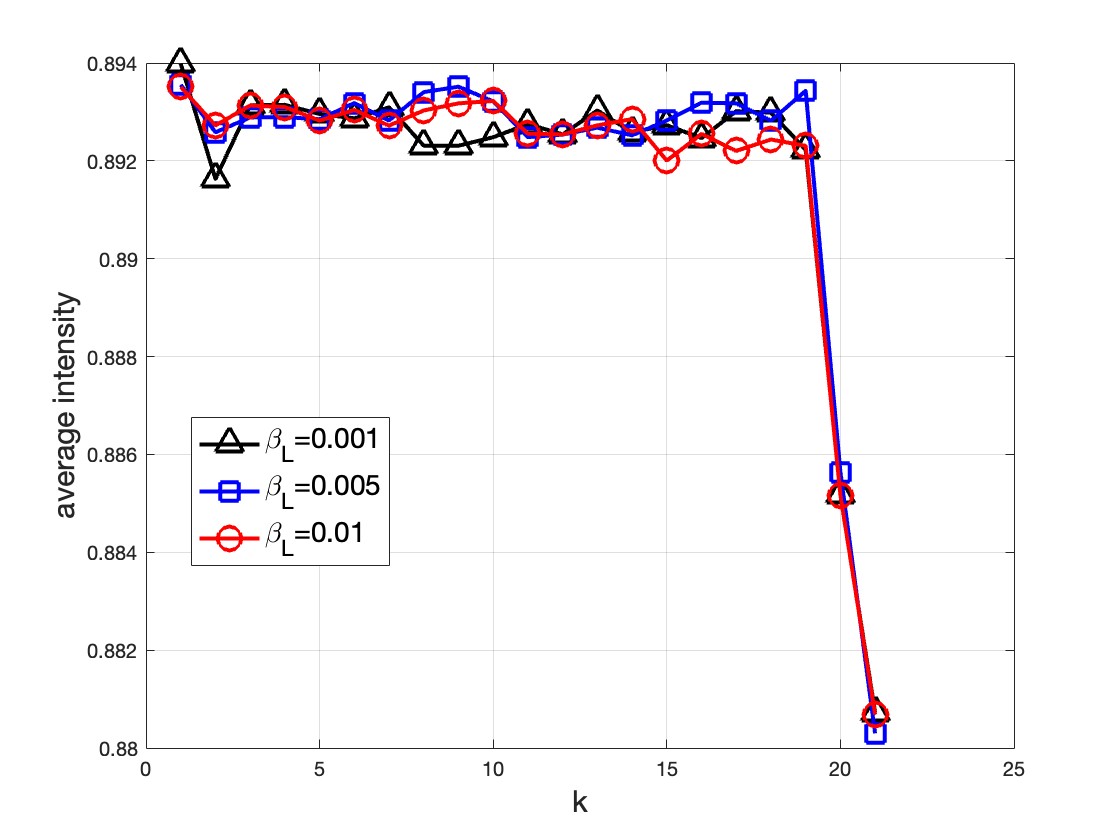}}
\hfill
{\includegraphics[width=0.49 \linewidth]{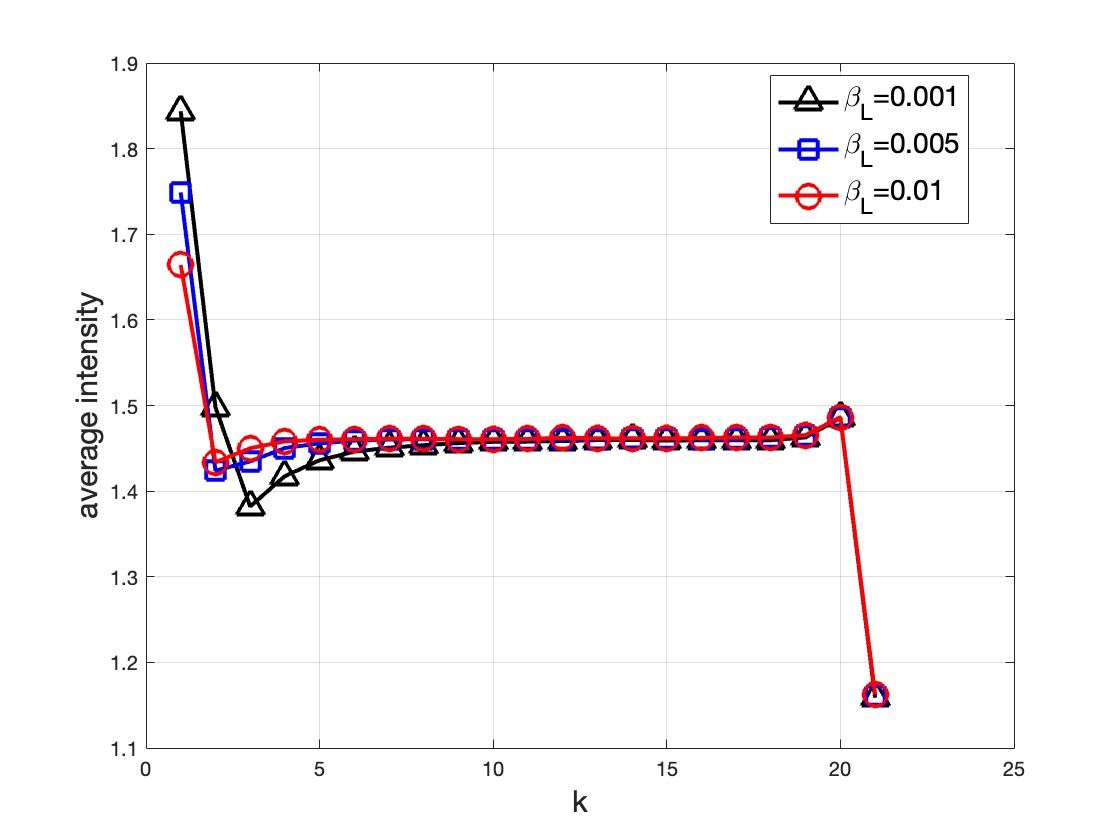}}
\vspace{-0.5cm}
\caption{Normalized average position (top row) and average intensity profiles (bottom row) profiles for the dynamics in Eq.~\eqref{eq:intens_1} (left panel) and in Eq.~\eqref{eq:intens_2} (right panel) with the $\beta$-FPUT potential, with $g_4=1$. Values of the other parameters are the same as in Fig.~\ref{fig:fig2}.}
  \label{fig:fig3}
\end{figure}

%\begin{figure}[htbp]
%\centering
%{\includegraphics[width=0.49 \linewidth]{Figs/fig2a-fpu-d1.pdf}}
%\hfill
%{\includegraphics[width=0.49 \linewidth]{Figs/fig2b-fpu-d1.pdf}}
%\caption{Average intensity profiles obtained with the FPU potential, using the Metropolis dynamics defined in Eq.~\eqref{eq:intens_1} (left panel) and its variant given in Eq.~\eqref{eq:intens_2} (right panel), in the non-homogeneous case with $\beta_L\in\{0.001,0.005,0.01\}$, $\beta_B = 0.03$ and $\beta_R = 0.1$, with $N = 21$, $a = 10$ and $\delta=1.5\ a$.
%Simulations are performed over $5\times10^7$ time steps.}
%  \label{fig:fig5}
%\end{figure}

\begin{figure}[t]
\centering
{\includegraphics[width=0.49 \linewidth]{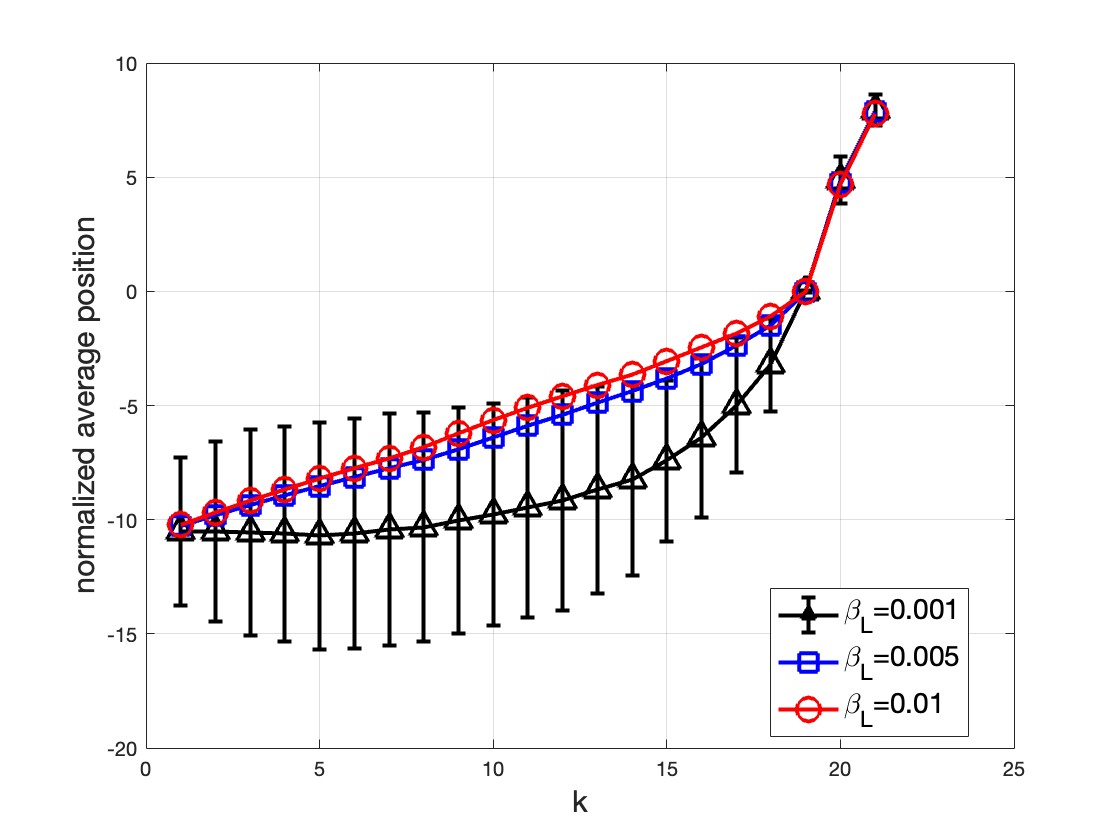}}
\hfill
{\includegraphics[width=0.49 \linewidth]{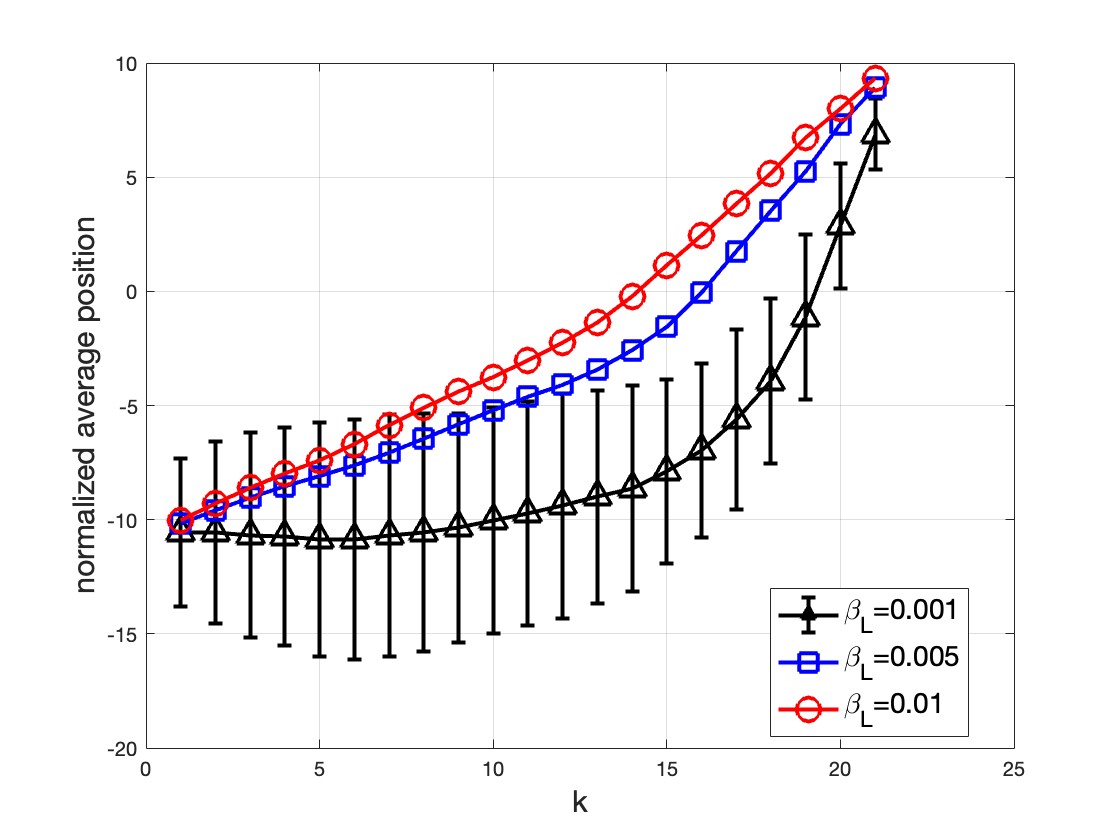}}
{\includegraphics[width=0.49 \linewidth]{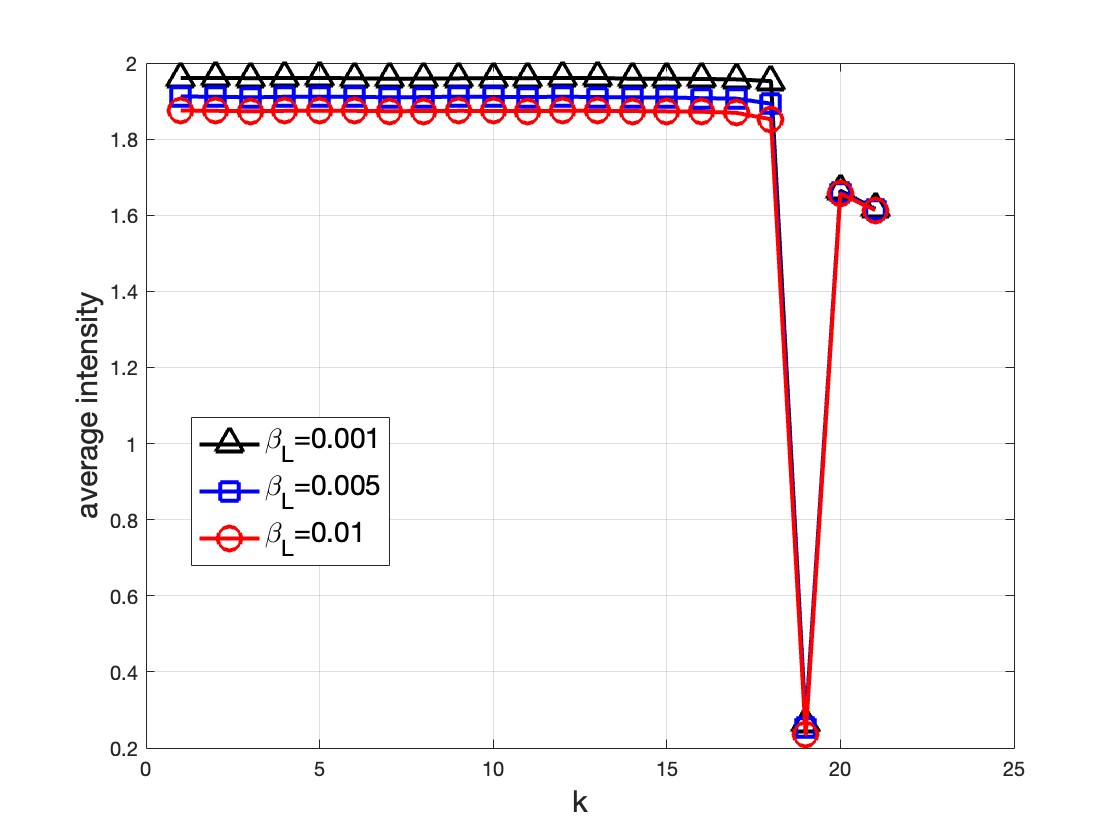}}
\hfill
{\includegraphics[width=0.49 \linewidth]{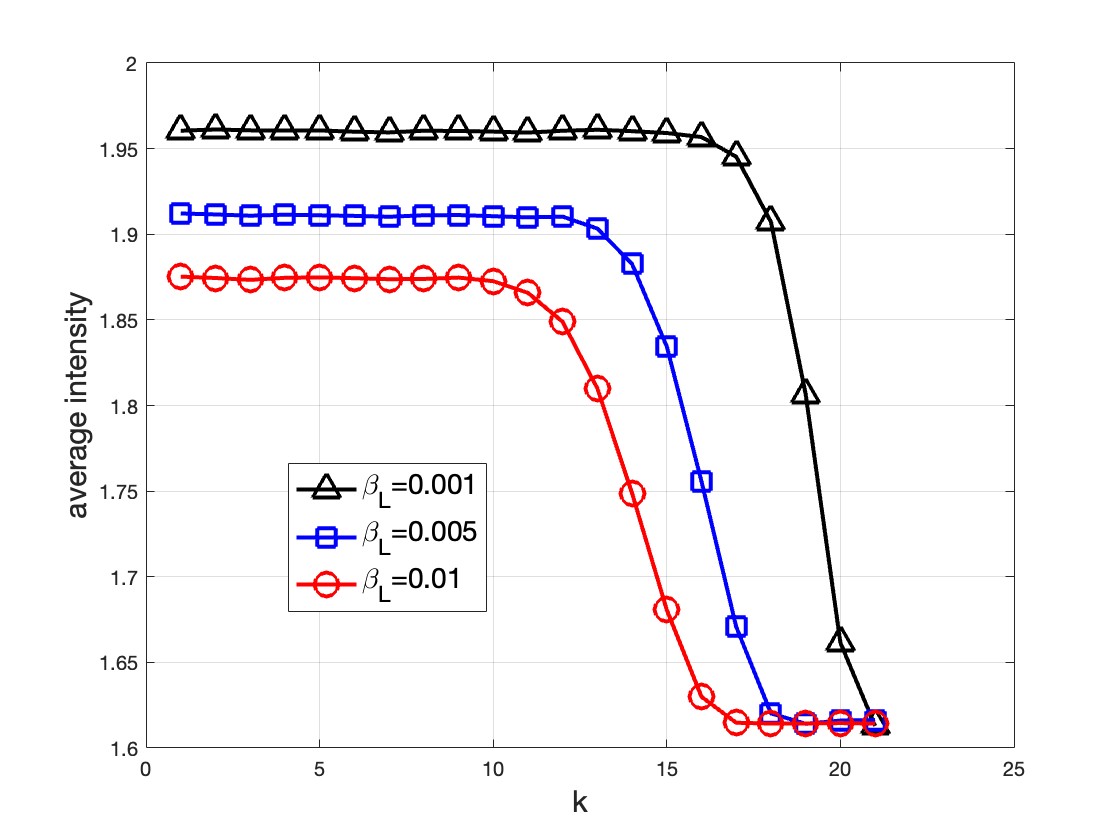}}
\vspace{-0.5cm}
\caption{Normalized average position (top row) and average intensity profiles (bottom row) profiles for the dynamics in Eq.~\eqref{eq:intens_1} (left panel) and in Eq.~\eqref{eq:intens_2} (right panel) with the harmonic potential, with $\delta=11\ a$. Values of the other parameters are the same as in Fig.~\ref{fig:fig2}.
%Normalized average position profiles obtained with the harmonic potential in Eq.~\eqref{harm}, using the Metropolis dynamics defined in Eq.~\eqref{eq:intens_1} (left panel) and its variant given in Eq.~\eqref{eq:intens_2} (right panel), in the non-homogeneous case with $\beta_L\in\{0.001,0.005,0.01\}$, $\beta_B = 0.03$ and $\beta_R = 0.1$, with $N = 21$, $a = 10$ and $\delta=11\ a$.
%Simulations are performed over $5\times10^7$ time steps.
\textcolor{black}{Note: the discontinuity in the bottom left panel is real.}
}
  \label{fig:fig4}
\end{figure}

%\begin{figure}
%\centering
%{\includegraphics[width=0.49 \linewidth]{Figs/fig2a-harm-d11.pdf}}
%\hfill
%{\includegraphics[width=0.49 \linewidth]{Figs/fig2b-harm-d11.pdf}}
%\caption{Average intensity profiles obtained with the harmonic potential using the Metropolis dynamics defined in Eq.~\eqref{eq:intens_1} (left panel) and its variant given in Eq.~\eqref{eq:intens_2} (right panel), in the non-homogeneous case with $\beta_L\in\{0.001,0.005,0.01\}$, $\beta_B = 0.03$ and $\beta_R = 0.1$, with $N = 21$, $a = 10$ and $\delta=11\ a$.
%Simulations are performed over $5\times10^7$ time steps.}
%  \label{fig:fig7}
%\end{figure}

\begin{figure}[t]
\centering
{\includegraphics[width=0.49 \linewidth]{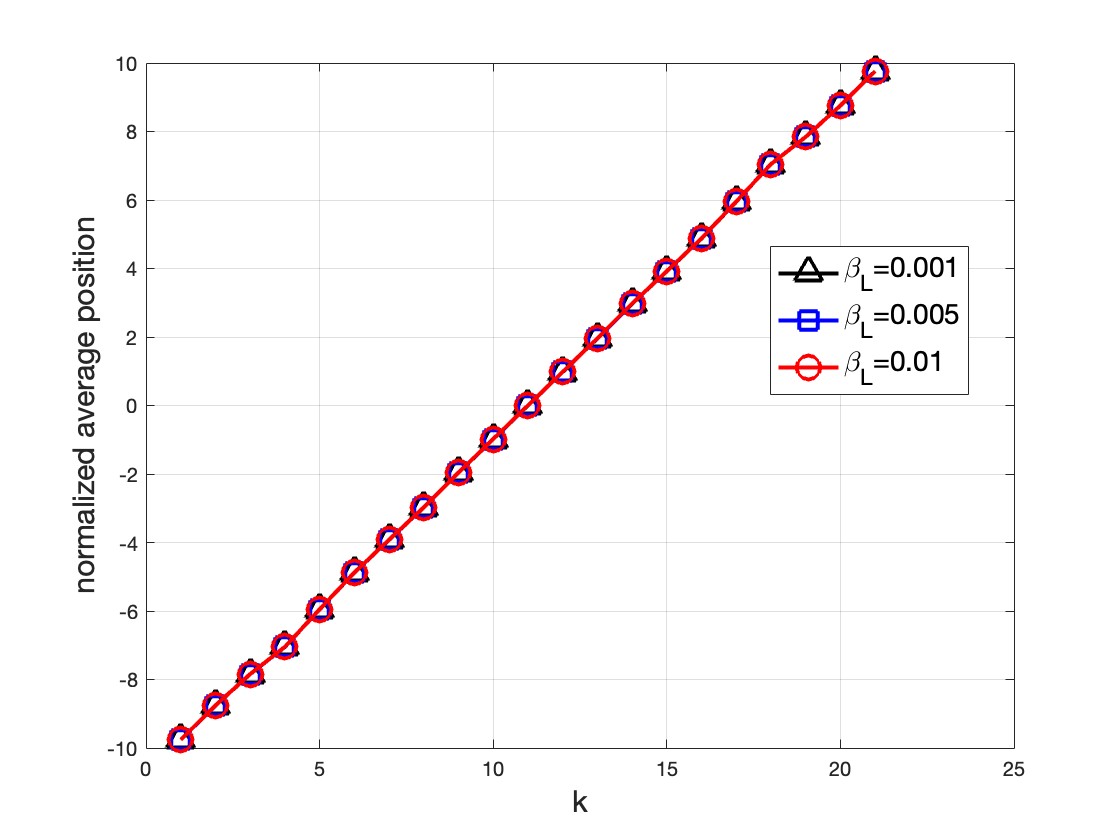}}
\hfill
{\includegraphics[width=0.49 \linewidth]{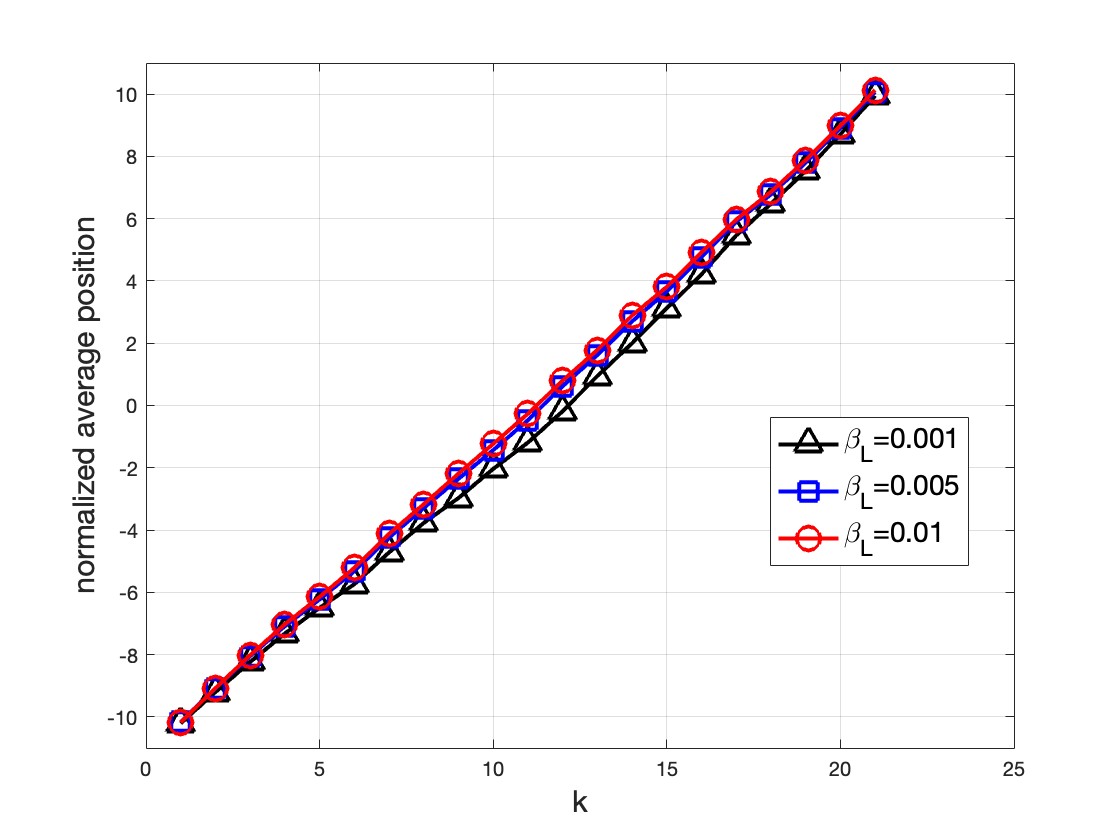}}
{\includegraphics[width=0.49 \linewidth]{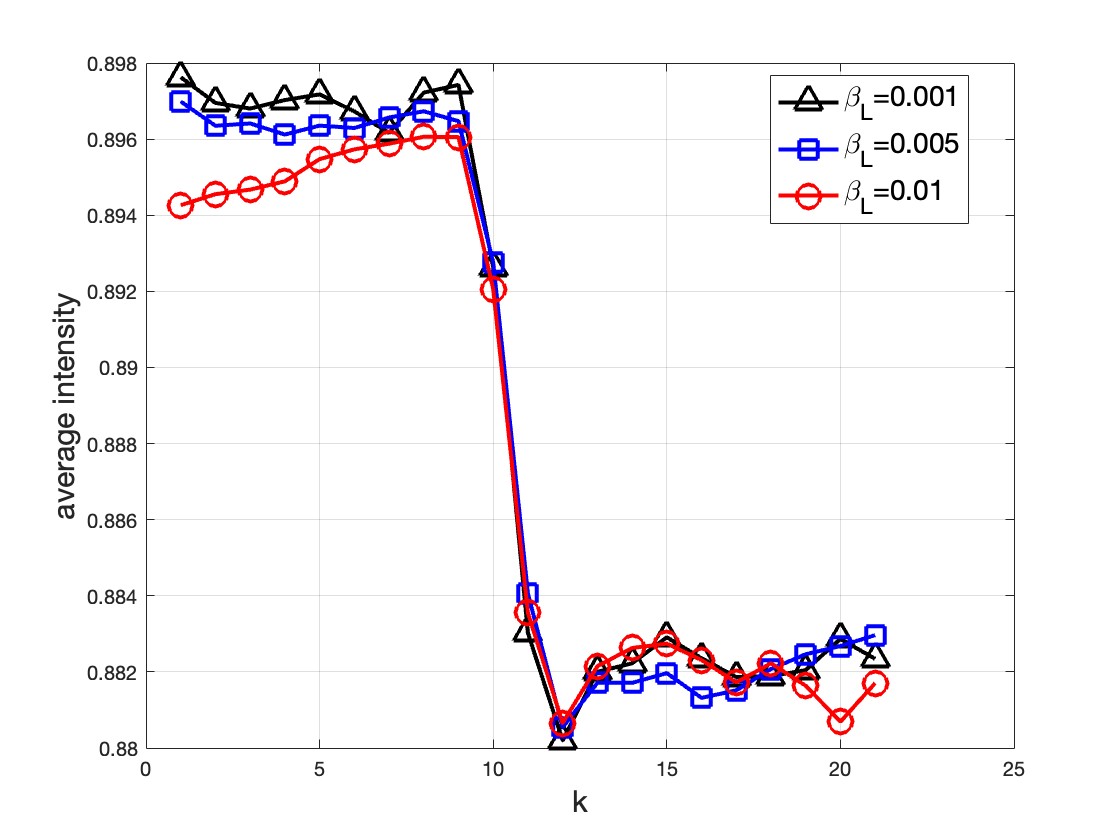}}
\hfill
{\includegraphics[width=0.49 \linewidth]{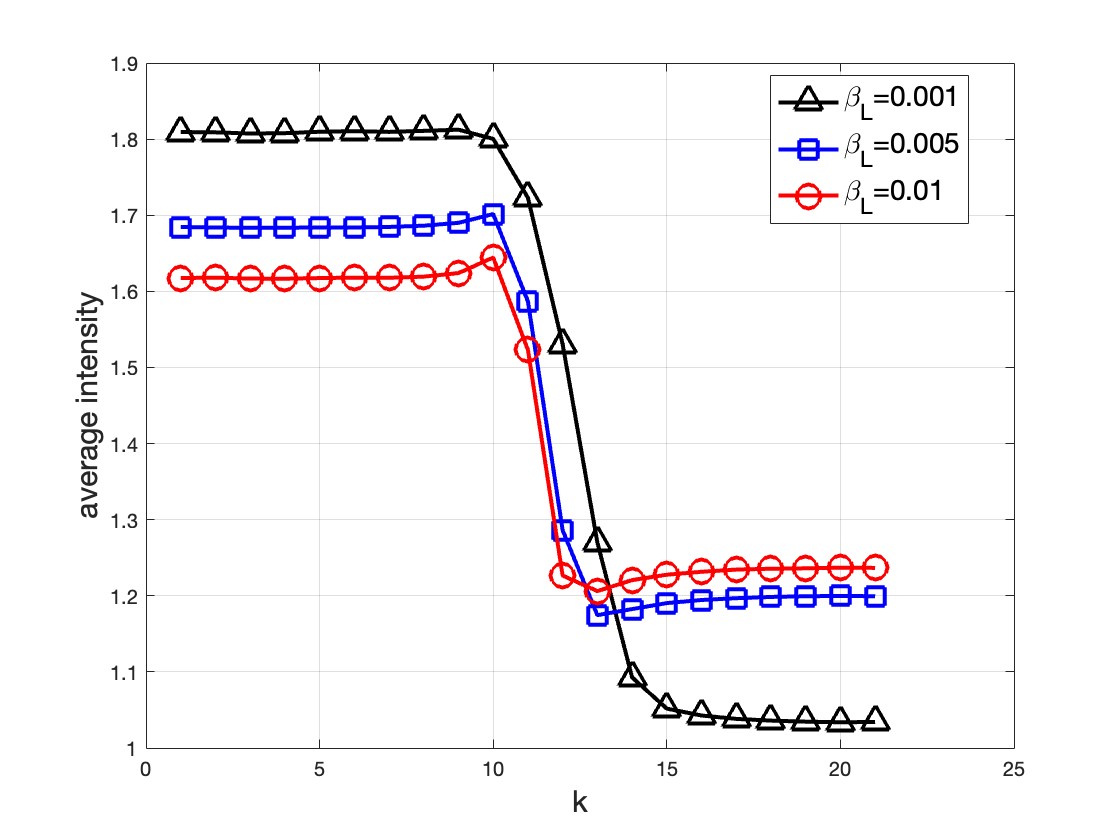}}
\vspace{-0.5cm}
\caption{Normalized average position (top row) and average intensity profiles (bottom row) for the dynamics in Eq.~\eqref{eq:intens_1} (left panel) and in Eq.~\eqref{eq:intens_2} (right panel) with the $\beta$-FPUT potential, with $\delta=11\ a$ and $g_4=1$. Values of the other parameters are the same as in Fig.~\ref{fig:fig2}.
%Normalized average position profiles obtained with the FPU potential in Eq.~\eqref{fpu}, using the Metropolis dynamics defined in Eq.~\eqref{eq:intens_1} (left panel) and its variant given in Eq.~\eqref{eq:intens_2} (right panel), in the non-homogeneous case with $\beta_L\in\{0.001,0.005,0.01\}$, $\beta_B = 0.03$ and $\beta_R = 0.1$, with $N = 21$, $a = 10$ and $\delta=11\ a$.Simulations are performed over $5\times10^7$ time steps.
}
  \label{fig:fig5}
\end{figure}

We present simulation results for $N=21$, 
$\beta_L=0.001, 0.005, 0.01$, $\beta_B=0.03$, $\beta_R=0.1$, 
and $a = 10$. In particular, the figures show the results for
the reversible and the irreversible dynamics, in the left and
right columns, respectively. The time-averaged positions of the
particles, $\langle x_k\rangle$, are shown in the top rows,
while the average intensities, $\langle c_k\rangle$, appear in
the bottom rows.
The number of time steps, in all simulations, is 
$5\times 10^7$, which is sufficient to reach the stationary state. 
Figure~\ref{fig:fig2} and Fig.~\ref{fig:fig4}
pertain to the harmonic case with $\delta=1.5\,a$ and
$\delta=11\,a$, whereas Fig.~\ref{fig:fig3} and
Fig.~\ref{fig:fig5} correspond to the $\beta$-FPUT potential for the
same values of $\delta$.
Our simulations reveal a sensible difference
between the particle position profiles for the two potentials.
In the $\beta$-FPUT case, the spatial profile of $\langle x_k\rangle$ 
is approximately independent of the dynamics and of the values
of $\beta_L$ and $\delta$, as it closely matches the lattice
profile $\bx^e$ in all cases; see the top rows in 
Fig.~\ref{fig:fig3} and Fig.~\ref{fig:fig5}. 

Conversely, for
the harmonic potential, under both dynamics the profiles of
$\langle x_k\rangle$ exhibit greater sensitivity to the inverse
temperature of the left reservoir, $\beta_L$, with deviations
from the linear trend of $\bx^e$ becoming increasingly marked
close to the left reservoir for smaller $\beta_L$ values.
Therefore, while in the 
$\beta$-FPUT potential the particles fluctuate
around the mechanical equilibrium configuration, in the
harmonic case the chain is strongly spatially inhomogeneous,
with a higher density near the hot thermal reservoir and,
more precisely, in a left neighborhood of the particle fixed at
$x_0$. This clustering occurs despite the
presence of large fluctuations of the particles around their
respective mean positions; see the error bars in
Fig.~\ref{fig:fig2} and Fig.~\ref{fig:fig4}. 
We also note
that there is a perfect coherence between the position and
intensity plots, since the average intensity is maximal for
those particles whose average position lies in the part of
the lattice with higher temperature.

%\cgr{[Questo comportamento bizzarro assomiglia a quello visto con le catene deterministiche con potenziale FPUT; dobbiamo citare questo fatto?]}

A similar scenario arises when the mean intensities 
$\langle c_k \rangle$ of the various models are compared. 
The results of the two dynamics with the $\beta$-FPUT 
potential yield qualitatively similar outcomes with respect 
to the intensities (see the bottom rows of Fig.~\ref{fig:fig3} 
and Fig.~\ref{fig:fig5}), while the harmonic case appears to 
depend more strongly on the dynamics (see the bottom rows of 
Fig.~\ref{fig:fig2} and Fig.~\ref{fig:fig4}). 
Moreover, in the $\beta$-FPUT case, 
%certain aspects are worth
%highlighting. 
the average intensity remains nearly constant
along most of the chain for small $\delta$ under both
dynamics; see Fig.~\ref{fig:fig3}. While with the
reversible dynamics this constant is close to the intensity of
the particles in the hot reservoir (left bottom panel), with
the irreversible dynamics it is approximately the mean of the
intensities of the particles closest to the two reservoirs
(right bottom panel).
This makes the intensity profile of the irreversible $\beta$-FPUT dynamics
%the intensity profile 
(Fig.~\ref{fig:fig3}, bottom right panel) 
\textcolor{black}{resemble that of temperature in harmonic chains \cite{rieder1967properties,giberti2011anomalies}, to which the microscopic state equation of \cite{giberti2011anomalies} applies.
Interestingly, 
%Regarding the role of temperature, it can be observed that 
small $\delta$ makes the mean intensity at the center of the chain} nearly independent of $\beta_L$, although different in the two dynamics, Fig.~\ref{fig:fig3}. The profile  more strongly depends on $\beta_L$ in the irreversible dynamics when $\delta$ is large,
as shown in the bottom right panel of Fig.~\ref{fig:fig5}.

%\cgr{[Resta da dire qualcosa su $\langle c_k\rangle $ nel caso armonico.???]}

%\begin{figure}[htbp]
%\centering
%{\includegraphics[width=0.49 \linewidth]{Figs/fig2a-fpu-d11.pdf}}
%\hfill
%{\includegraphics[width=0.49 \linewidth]{Figs/fig2b-fpu-d11.pdf}}
%\caption{Average intensity profiles obtained with the FPUT potential, using the Metropolis dynamics defined in Eq.~\eqref{eq:intens_1} (left panel) and its variant given in Eq.~\eqref{eq:intens_2} (right panel), in the non-homogeneous case with $\beta_L\in\{0.001,0.005,0.01\}$, $\beta_B = 0.03$ and $\beta_R = 0.1$, with $N = 21$, $a = 10$ and $\delta=11\ a$.
%Simulations are performed over $5\times10^7$ time steps.}
%  \label{fig:fig9}
%\end{figure}

\section{Conclusions}
\label{sec:sec3}

In this work we have studied models of \textcolor{black}{stochastic chains} within the framework of interacting particle systems. In our set-up particles hop on a discrete lattice corresponding to the set $\mathbb{Z}$ of integer numbers and are subject to local interactions in index space. An assigned inverse temperature profile, corresponding to a three steps function, influences particle jumps, thus mimicking the coupling with external thermal reservoirs at different temperatures. 
We analyzed the stationary profiles of both particle positions and jump intensities in systems featuring either a harmonic or a Fermi-Pasta-Ulam-Tsingou potential. The particle dynamics were considered in two variants. The first, given by Eq.~\eqref{eq:intens_1}, follows the Metropolis algorithm and is reversible with respect to a stationary distribution that retains the structure of the Gibbs measure for equilibrium systems. 
The second, defined by Eq.~\eqref{eq:intens_2}, instead breaks detailed balance 
\textcolor{black}{at the interfaces between two regions of the lattice}
but, similarly to the first, exhibits notable stationary effects. In particular, when particles interact through a harmonic potential, the irreversible dynamics leads to a higher concentration of particles near the hotter reservoir, a phenomenon akin to that observed in certain non-equilibrium $\beta$-FPUT chains
\cite{di2024microscopic}.
\textcolor{black}{This is analogous to the case of stretched oscillators chains.}
The analytical investigation of stationary profiles for inhomogeneous models will be discussed in a future work.

\vspace*{1cm}

\begin{acknowledgements}
This research was performed under the auspices of Italian National Group of Mathematical
Physics (GNFM) of the National Institute for Advanced Mathematics - INdAM.
ENMC and MC thank the PRIN 2022 project
``Mathematical Modelling of Heterogeneous Systems (MMHS)'',
financed by the European Union - Next Generation EU,
CUP B53D23009360006, Project Code 2022MKB7MM, PNRR M4.C2.1.1.
LR acknowledges support from NODES Program,  MUR - M4C2 1.5 PNRR, funded by the European Union - NextGenerationEU (Grant agreement no. ECS 00000036 ). 
MC thanks Stefano Lepri (Istituto dei Sistemi Complessi, CNR, Italy) for useful discussions.

\end{acknowledgements}

\bibliographystyle{abbrv}
\bibliography{biblio}

%\bibliographystyle{...}
%\bibliography{...}
\end{document}